\pdfoutput=1
\documentclass[11pt]{article}

\usepackage[]{EMNLP2022}

\usepackage{times}
\usepackage{latexsym}

\usepackage[T1]{fontenc}
\usepackage[utf8]{inputenc}

\usepackage{microtype}

\usepackage{inconsolata}

\usepackage{cite}
\usepackage[T1]{fontenc}    % use 8-bit T1 fonts
\usepackage{hyperref}       % hyperlinks
\usepackage{url}            % simple URL typesetting
\usepackage{booktabs}       % professional-quality tables
\usepackage{amsfonts}       % blackboard math symbols
\usepackage{nicefrac}       % compact symbols for 1/2, etc.
\usepackage{microtype}      % microtypography
\usepackage{amsmath}
\usepackage{savesym}
\savesymbol{checkmark}
\usepackage{lipsum,multicol}
\usepackage{graphicx}
\usepackage{xcolor}
\usepackage{hyphenat}
\def\eqref#1{equation~\ref{#1}}
\def\1{\bm{1}}

\def\rvb{{\mathbf{b}}}
\def\rvc{{\mathbf{c}}}

\def\rvh{{\mathbf{h}}}

\def\rvp{{\mathbf{p}}}
\def\rvq{{\mathbf{q}}}

\def\rvx{{\mathbf{x}}}

\def\rmW{{\mathbf{W}}}

\DeclareMathAlphabet{\mathsfit}{\encodingdefault}{\sfdefault}{m}{sl}
\SetMathAlphabet{\mathsfit}{bold}{\encodingdefault}{\sfdefault}{bx}{n}

\newcommand{\sigmoid}{\sigma}

\usepackage{dingbat}
\usepackage{enumerate}
\usepackage{subfig}
\usepackage{natbib}

\usepackage[symbol]{footmisc}
\usepackage{booktabs}
\usepackage{multirow}
\usepackage{adjustbox}

\newcommand{\ml}[1]{\textcolor{orange}{\bf [ml: #1]}}

\title{GPT-based Open-Ended Knowledge Tracing for Computer Science Education}

\author{Naiming Liu$^{*}$, Zichao Wang$^{*}$, Richard G. Baraniuk \\
  Rice University \\
  \texttt{nl35,zw16,richb@rice.edu} \\\And
  Andrew Lan \\
  University of Massachusetts Amherst \\
  \texttt{andrewlan@cs.umass.edu} \\}

\begin{document}
\maketitle
\begin{abstract}
% \footnote[1]{text}
{\renewcommand{\thefootnote}{\fnsymbol{footnote}}\footnotetext[1]{The first two authors contributed equally.}}
In education applications, {\em knowledge tracing} refers to the problem of estimating students' time-varying concept/skill mastery level from their past responses to questions and predicting their future performance.
One key limitation of most existing knowledge tracing methods is that they treat student responses to questions as \emph{binary-valued}, i.e., whether they are correct or incorrect. 
Response correctness analysis/prediction ignores important information on student knowledge contained in the exact content of the responses, especially for open-ended questions.
In this paper, we conduct the first exploration into {\em open-ended knowledge tracing} (OKT) by studying the new task of predicting students' exact open-ended responses to questions.
Our work is grounded in the domain of computer science education with programming questions. 
We develop an initial solution to the OKT problem, a student knowledge-guided code generation approach, that combines program synthesis methods using language models with student knowledge tracing methods. 
We also conduct a series of quantitative and qualitative experiments on a real-world student code dataset to validate OKT and demonstrate its promise in educational applications.
\end{abstract}

\section{Introduction}
\renewcommand{\thefootnote}{\arabic{footnote}}

Knowledge tracing (KT) \citep{kt} refers to the problem of estimating student mastery of concepts/skills/knowledge components from their responses to questions and using these estimates to predict their future performance.
KT methods play a key role in many of today's
%have been applied in today's 
large-scale online learning platforms to automatically estimate the knowledge levels of a large number of students and provide each of them with personalized feedback and recommendation, leading to improved learning outcomes \citep{ritter}.
KT methods consist of two essential components. 
First, a \emph{knowledge estimation} (KE) component, i.e., 
\begin{align}
\label{eq:ku}
    \rvh_{t+1} = \text{KE} ((\rvp_1,\rvx_1), \ldots, (\rvp_{m},\rvx_{t})),
\end{align}
estimates a student's current knowledge state $\rvh_{t+1}$ using questions ($\rvp$) and responses ($\rvx$) from previous (discrete) time steps for this student. 
Second, a \emph{response prediction} (RP) component predicts the student's response to the next question (or future questions), i.e., $\rvx_{t+1} \sim \text{RP} (\rvh_{t+1},\rvp_{t+1})$.
Section~\ref{sec:rw} contains a detailed overview of existing KT methods and how the question, responses, and knowledge state variables are represented. 

One key limitation of almost all existing KT methods is that they only analyze and predict \emph{binary-valued} student responses to questions, i.e., the \emph{correctness} of the response.
That is, the RP is typically a simple binary classifier. As a result, one can broadly apply KT methods to any question as long as student responses are graded. However, this approach loses important information regarding student mastery, since it does not make use of the \emph{content} of questions and student responses, especially for open-ended questions. Past work has shown that students' open-ended responses to such questions contain useful information on their knowledge states, e.g., having a ``buggy rule'' \citep{brown}, exhibiting misconceptions \citep{feldman,junchen,smith}, or a general lack of knowledge \citep{anderson}; this information is highly salient for instructors but cannot be captured by response correctness alone. 

%This limitation may prevent KT methods from extracting knowledge estimates that reflect specific errors and making personalized recommendations that help students correct these errors. 

Generative language models such as GPT \citep{gpt} provide an opportunity to fully exploit the rich information contained in open-ended student responses in various domains for the purposes of KT. In this paper, we focus on computer science education, where short programming questions require students to write code chunks that satisfy the question's requirements. The program synthesis capabilities of variants of pre-trained neural language models such as CodeX \citep{codex} enable the generation of short chunks of code from natural language instructions, which we can leverage for open-ended response prediction. 
However, two key challenges make this task difficult: 
First, as students learn through practice, their knowledge on different programming concepts is \emph{dynamic}; students can often learn and correct their errors given instructor-provided feedback or even error messages generated by the compiler. Therefore, {\em we need new KE models that can effectively trace time-varying student knowledge throughout their learning process.} 
Second, student-generated code is often \emph{incorrect and exhibits various errors}; there may also exist multiple correct responses that capture different lines of thinking among students. This intricacy is not covered by program synthesis models, since their goal is to generate correct code and they are usually trained on code written by skilled programmers. Therefore, {\em we need new RP models that can generate student-written (possibly erroneous) code that reflects their (often imperfect) knowledge of programming concepts.} 

%For grade-school level math problems, language models fine-tuned on mathematical content can generate short solutions consisting of a mixture of text and mathematical expressions \citep{gsm8k,math}. 

\subsection{Contributions}
%\noindent\textbf{Contributions:} 
In this paper, we present the first attempt at analyzing and predicting exact, open-ended student responses, specifically for programming questions in computer science education. {\renewcommand{\thefootnote}{\fnsymbol{footnote}}\footnotetext[1]{Find our code at \url{https://github.com/lucy66666/OKT}}} Our contributions can be summarized as follows:
%\vspace{-.5cm}
\begin{itemize}
\itemsep0em
    \item We define the {\bf open-ended knowledge tracing (OKT)} framework, a novel KT framework for open-ended student responses, and a new KT task, exact student response prediction. We ground OKT in the domain of computer science education for student code submission analysis and prediction but emphasize that \emph{OKT can be broadly applicable to a wide range of subjects that involve open-ended questions}.
    \item We develop an initial solution to the OKT task, a {\bf knowledge-guided} code generation method. Our method combines KE components in existing binary-valued KT methods with code generation models, casting the OKT task as a \emph{dynamic controllable generation} problem where the control, i.e., time-varying student knowledge states, are also learned. 
    %We also show that OKT is compatible with several popular KT methods that were originally developed for binary-valued student responses. 
    %We hope that OKT will open up a new direction in KT research that leverages generative content models. 
%    \item Third, we define a series of evaluation metrics for open-ended programming questions, ranging from metrics on code prediction accuracy, including exact match and partial match leveraging the inherent tree structure of code, to metrics on how the estimated knowledge states correlate with actual student-generated code. 
    \item Through {\bf extensive experiments on a real-world student code dataset}, we explore the effectiveness of OKT in reflecting variations of student code and especially errors in its knowledge state estimates. We explore the effectiveness of our solution in making reasonably accurate predictions of student-submitted code. We also discuss how these OKT capabilities can help computer science instructors and outline several new research directions. 
    %Our key takeaways are: ... you can do full response prediction but only to a certain extent depending on how much variation there is in student responses, better generative models help, there is some variation in actual student code that you can explain with knowledge states, etc. 
    %We also discuss the limitations of OKT and propose several important directions for future work.
\end{itemize}

\section{OKT for Computer Science Education}

\label{sec:cse}
\begin{figure}[t]
    \centering
    %\vspace{-2pt}
    \includegraphics[width=\linewidth]{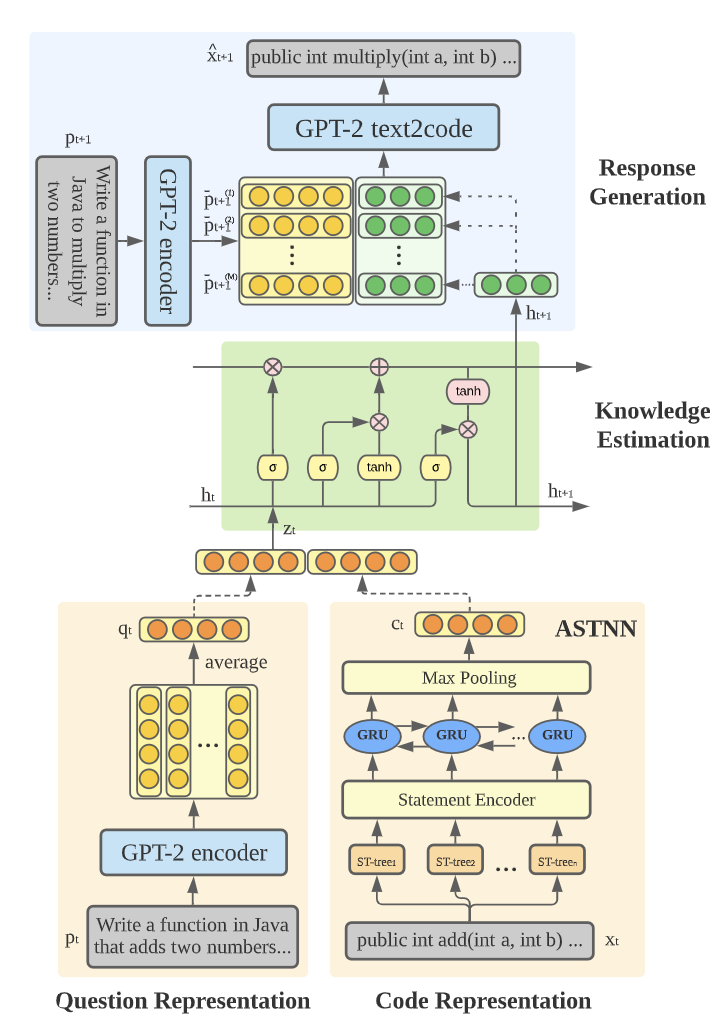}
    \vspace{-12pt}
    \caption{
    Open-ended knowledge tracing (OKT) block diagram.  We update the student's current knowledge state $\rvh_{t+1}$ using the last question $\rvp_t$ and actual student code $\rvx_t$. We then combine it with the next question statement $\rvp_{t+1}$ to generate our prediction of the actual student code $\hat{\rvx}_{t+1}$.}
    \vspace{-15pt}
    \label{fig:okt-code}
\end{figure}

%\ml{note for unifying terminology: domain should be CS Ed, where we have programming questions, students submit code as response to these questions}
We now define the OKT framework and detail specific model design choices in the domain of computer science education, where we focus on analyzing students' code submissions to programming questions. 
Figure~\ref{fig:okt-code} illustrates the three key components of OKT: knowledge representation (KR), KE, and response generation (RG), the last of which is the key difference between OKT and existing KT methods. 
Our key technical challenges are (i) how to represent programming questions and student code submissions (KR, Section~\ref{sec:kr}) and use them to estimate student knowledge states (KE, Section~\ref{sec:kt}); (ii) how to combine knowledge states with the question prompt to generate student code (RG, Section~\ref{sec:gen-model}); and (iii) how to efficiently perform optimization to train the OKT model components (Section~\ref{sec:loss}).
%Next, we address the above challenges and describe our design choices for each OKT component for CS education. 

%\vspace{-2pt}
\subsection{Knowledge Representation (KR)}
\label{sec:kr}

The purpose of the KR component is to convert the prompt/statement of questions that students respond to and their corresponding code submissions to continuous representations. Our KR component is significantly different from existing binary-valued KT methods that ignore question/response content and one-hot encode them using question/concept IDs and response correctness. %; see Section~\ref{sec:rw} for details. 

\noindent\textbf{Question Representation:}~
%Question prompts in programming assignments are usually in natural language. We thus use a pre-trained neural language model to convert them into embedding vectors. 
%In order to be consistent with the code generative model, 
We adopt the popular GPT-2 model\footnote{One can use any language model; we choose GPT-2 since our RG component for student code is also built on GPT-2.} \citep{gpt-2} for prompt representation: Given a question prompt $\rvp$, GPT-2 tokenizes it into a sequence of $M$ word tokens, where each token has an embedding $\bar{\rvp}_m \in \mathbb{R}^K$. For GPT-2, the dimension of these embeddings is $K = 768$. This procedure produces a sequence of token embeddings $\{\bar{\rvp}_1, \bar{\rvp}_2, \ldots, \bar{\rvp}_M\}$. 
%\in \mathbb{R}^{M \times K}$. 
We then average the embeddings of each prompt token to get our prompt embedding $\rvq = \sum_{m=1}^M \frac{\bar{\rvp}_{m}}{M}$, where the average is computed element-wise on vectors. 
%[\bar{e}_1, \bar{e}_2, \ldots, \bar{e}_K] \in \mathbb{R}^{K}$, where $\bar{e}_k = \sum_{i=1}^M \frac{e_{ik}}{M}$.  Then we use the prompt representation $q$ as the input to our knowledge tracing model. 
%\ml{so we're like simply averaging the gpt token embeddings of each prompt token; is this how people usually encode things with gpt?} \nl{not sure exactly, will look into it} \zw{can say that we can opt to use other types of encoding methods. also did we use gpt or bert? i don't remember} \\

\noindent\textbf{Code Representation:}~
In order to preserve both semantic and syntactic properties of programming code in embedding vectors, we utilize ASTNN \citep{astnn}, a popular tool for code representation. We first parse student-submitted code into an abstract syntax tree (AST). We then split each full AST into a sequence of non-overlapping statement trees (ST-trees) through preorder traversal. 
Each ST-tree contains a statement node as the root and its corresponding AST nodes as children. 
We then pass the ST-trees through a recurrent statement encoder
%, which contains an embedding layer using pre-trained Word2Vec and a RNN based encoder layer.
%After extracting the statement vector representations of ST-trees, 
to obtain embedding vectors and use a bidirectional gated recurrent unit network \citep{gru} to capture the naturalness of the statements and further enhance the capability of the recurrent layer. Eventually, we apply a max-pooling layer to capture the most important semantics for each dimension of the embedding. We denote this entire process as 
%\begin{equation*}
$\rvc = \text{ASTNN} (\rvx)$
%\end{equation*}
where $\rvx$ is student-submitted code and $\rvc$ is its code embedding vector, which we use as input to the KE component. 
We refer readers to \citep{astnn} for more details on ASTNN. 

\subsection{Knowledge Estimation (KE)}
\label{sec:kt}
% We use the most straight-forward knowledge tracing model DKT for illustration.
The purpose of the KE component is to turn a student's past question/code information into estimates of their current knowledge state. 
Following DKT \citep{dkt}, a popular existing KT method, we use a long short-term memory (LSTM) model \citep{lstm} to update a student's current knowledge state, $\rvh_{t+1}$, given their previous response at the last time step. 
%\ml{can cut details on LSTM if necessary - probably safe to assume readers would know that} 
% LSTMs have three major parts: the forget gate, the input gate and the output gate. The forget gate removes probably irrelevant contents from the cell state. The input gate first calculates a candidate vector and then a weight vector (ranging from 0 to 1) to determine the proportion of the candidate vector to add to the current memory state. Finally the output gate proportionally combines the knowledge gained from previous gates as the hidden states $\rvh_t \in \mathbb{R}^D$.
%
%Instead of using one-hot embeddings for questions and responses according to question/concept IDs, 
We use the output of the KR component, i.e., question prompt and code embeddings, as the input to the KE component as
%As shown in Eq.~\ref{eq:2}, the input to the model is a time series which consists of the embeddings of question prompts and student-submitted codes, denoted as $Z = \{(\rvq_t, \rvc_t)\}_{t=1}^T$, where $z_t = (\rvq_t, \rvc_t)$. 
%Our KE component can be summarized as
%\begin{align*}
$\rvh_{t+1} = \text{LSTM}(\rvh_{t}, \rvq_t, \rvc_t)$
%\end{align*}
%where $\rh_t$ is the student's knowledge state at time step $t$. 
and use it as input to the RG component to generate predicted student code submissions. 
In principle, we can use any existing binary-valued KT method as OKT's KE component. We validate in our experiments (Section~\ref{sec:exp}) that OKT is compatible with two other popular KT methods, DKVMN \citep{dkvmn} and AKT \citep{akt}, that are based on external memory and attention networks. 

\subsection{Response Generation (RG)}
\label{sec:gen-model}
The purpose of RG, OKT's core component, is to \textbf{predict open-ended responses}, i.e., generate predicted student code, which makes OKT significantly different from existing binary-valued KT methods with binary classifiers of response correctness.
%which aims to predict the student written codes from problem prompts and their corresponding knowledge states. 
We fine-tune a base GPT-2 generative model into a text-to-code model $P_\Theta$ with parameter $\Theta$ on code data (see Section~\ref{sec:pre-train} for details). We choose language models over other code generation approaches since their text-to-code generation pipeline suits OKT well. %We detail the pre-training procedure in Section~.

Our key technical challenge is how to use knowledge states as \emph{control} in the code generation model to guide personalized code predictions for each student. 
%Take a single student as an example, at time step $t$, we can get a knowledge state $h_t$ from the knowledge tracing model which indicates the relevant knowledge the student has already obtained. Then we choose the question prompt $p_{t+1}$ which should be given to the student at the following time step $t+1$. 
Given the current question prompt, $\rvp_{t+1}$, and its sequence of $M$ token embeddings $\{\bar{\rvp}_1, \bar{\rvp}_2, \ldots, \bar{\rvp}_M\}$, where $\bar{\rvp}_m \in \mathbb{R}^K$ (we drop the time step index $t$ in prompt tokens for clarity), our approach for injecting student knowledge states into the code generation model is to replace raw token embeddings with \emph{knowledge-guided} embeddings using an \emph{alignment} function, i.e., %Since we need the code generative model to take student's previous knowledge into consideration, we use a function to combine every token embeddings and knowledge states together, denoted as 
$\rvp_{m} = f(\bar{\rvp}_{m}, \rvh_{t+1})$ for $m = 1, \ldots, M$. Therefore, the GPT-2 input embeddings are
\begin{equation*}
    \{\rvp_{1}, \ldots, \rvp_{M}\!\} \!=\! \{\!f(\bar{\rvp}_1, \rvh_{t+1}), \ldots, f(\bar{\rvp}_M, \rvh_{t+1})\!\}. 
\end{equation*}
%where $E_{t}$ is the set of all token embeddings to the code generation model at time step $t$. 
Intuitively, this (possibly learnable) alignment function aligns the space of knowledge states with the space of textual embeddings for the question prompt. Thus, knowledge states are responsible for predicting different code submitted to the same programming question by different students. 
%There exists a time step mismatch between the question prompt $p_{t+1}$ and knowledge state $h_t$ because we need the students' previous knowledge to predict on the following unseen questions. 

We explore four different alignment functions to combine knowledge states with question prompt token embeddings:
\begin{itemize}
\itemsep0em
    \item Addition, i.e., $\rvp_{m} = \bar{\rvp}_{m} + \rvh_{t+1}$.
    \item Averaging, i.e., $\rvp_{m} = (\bar{\rvp}_{m} + \rvh_{t+1})/2$.
    \item Weighted addition, i.e., using a learnable weight for knowledge states, $\rvp_{m} = \bar{\rvp}_{m} + \alpha \cdot \rvh_{t+1}$.
    \item Linear combination, i.e., applying a learnable affine transformation to the knowledge states before adding it to token embeddings, $\rvp_{m} = \bar{\rvp}_{m} + \mathbf{A} \rvh_{t+1} + \mathbf{b}$.
\end{itemize}
The latter two functions are learnable with parameters $\alpha \in \mathbb{R}$, $\mathbf{A} \in \mathbb{R}^{D \times K}$, and $\rvb \in \mathbb{R}^K$. 
%We will remove the time step index $t$ during the description of the generative process below. 
Therefore, the predicted student code (with $N$ code tokens), $\rvx= \{x_1, x_2, \ldots, x_N\}$, is generated in an autoregressive manner by the RP component given the knowledge-guided question prompt token embeddings $\{\rvp_1, \ldots, \rvp_M\}$. %

%the autoregressive generation model for student code $\rvx = \{x^{(1)}, x^{(2)}, \ldots, x^{(N)}\}$ (we remove the time step index $t$ for now for clarity and $N$ denotes the number of code tokens) given the knowledge-guided question prompt token embeddings $\{\rvp^{(1)}, \ldots, \rvp^{(M)}\}$ can be written as
%\begin{align}
%    \rvx & \sim \textstyle \prod_{j=1}^N P_\Theta \big(x^{(j)} | \{\rvp^{(1)}, \ldots, \rvp^{(M)}\}, \notag \\ & \quad \quad  \{x^{(j')}\}_{j'=1}^{j-1} \big), \label{eq:lm}
%\end{align}
%\begin{equation*}
%    x \sim P_\Theta (x | \{\rvp^{(1)}, \ldots, \rvp^{(M)}\}), %\label{eq:gen}
%\end{equation*}

%Since GPT-2 is an autoregressive language model, we can further decompose Eq.~\ref{eq:gen} into 

%where $x^{(j)}$ denotes the $j^\text{th}$ code token. 

\subsection{Optimization}
\label{sec:loss}
During the training process, we jointly optimize the parameters of the KE and RG components of OKT; in essense, we are learning \emph{both} a controllable generation model for student responses \emph{and} the control itself, which is the student's time-varying knowledge state. We keep the knowledge representation encoders $E_1$ and $E_2$ fixed. The objective for one student code submission is given by %simply the negative log-likelihood/cross entropy loss for all code tokens:
\begin{align}
    \notag 
    {\sl Loss} & = \textstyle \sum_{n=1}^N - \log P_\Theta \big( x_n | \; \{\rvp_1, \ldots, \rvp_M\}, \\
    & \quad \quad \{x_{n'}\}_{n'=1}^{n-1} \big), \label{eq:loss-gpt}
\end{align}
where $\Theta$ denotes the set of parameters in the RP component, including both the GPT-2 text-to-code model parameters and learnable parameters in the alignment function $f(\cdot)$. The final training objective is the sum of this loss over all code submissions made by all students.
%where we have decomposed the loss function for a single loss across tokens due to GPT's autoregressive nature, while the  
%The above training objective serves as a proxy that optimizes for code quality. It was minimized using stochastic gradient descent on minibatches. 

%%%
% this can be cut (i mean move to supp) if we need space :(
We also design an efficient training setup for OKT. For existing neural network-based KT methods, at each training step, we use a batch of student (question, response) sequences to compute the correctness prediction loss across all time steps and all students in the batch. 
We cannot use this training method since OKT's loss for one student is the sum of code prediction losses over all time steps, whereas the loss at each time step is itself the sum of a sequence of cross entropy losses for code token predictions. 
As a result, if we use the training setup for existing KT methods, at each training step, we need to call the response generator for a total of $T\times B$ times where $T$ is the number of time steps and $B$ is the batch size, which will significantly slow down training. 
Instead of batching over students, we use a batch of (student, time step) pairs. Then, at each training step, we first apply the knowledge update component in OKT to compute the knowledge states for students in the batch, extract the knowledge states corresponding to the sampled time steps in the batch, and then feed them into the response generator. This setup enables efficient training for OKT.
%This training setup is more akin to the training setup for language models than that for KT methods, with an added, time-varying KT component, and is $T$ times more efficient than the KT training setup. 
%\zw{terminology needs unification}
%\zw{my TODO: draw a plot} 
%\ml{low priority i think - i removed the last sentence since the classic KT training setup is the same as LM training, right?}\zw{yes - got that}

\subsection{Pre-training Models}
\label{sec:pre-train}
%\noindent \textbf{Knowledge Tracing Models:} 
Before training OKT, we pre-train its KE component using the binary-valued correctness prediction loss with question and code embeddings as input, following \citep{ncsupkt,pkt}. 
%\noindent \textbf{Text-to-code Models:} 
%We also pre-train the response generation component to adapt GPT-2 to the text-to-code task with actual code. 
Since we cannot directly use CodeX \citep{codex} due to our need to adjust the input embeddings with student knowledge states, we pre-train a text-to-code pipeline by fine-tuning a standard GPT-2 model on the Funcom dataset \citep{funcom}, which contains 2.1 million Java code snippets and their textual descriptions. 

\section{Experiments}
\label{sec:exp}

We now present a series of experiments to explore the capabilities of OKT. 
We first introduce the dataset, various quantitative metrics on which we evaluate various methods, and detail quantitative results. 
%Since OKT is a new task without benchmark baselines to compare with, we define several variants of our proposed OKT method to demonstrate its applicability and compatibility with several representative knowledge tracing methods and code generation models. 
%and (ii) how robust is our method with such changes in the underlying components. \ml{ii (above) is a bit weird - maybe cut?} 
We then qualitatively illustrate that OKT (i) learns a meaningful latent student knowledge space and (ii) generates predicted student code that
%, although not always exactly matching the students' solutions, 
capture their coding patterns and error types. 

%\vspace{5pt}
\noindent\textbf{Dataset:}~
We use the dataset from the CSEDM Data Challenge, henceforth referred to as the {\bf CSEDM dataset}.\footnote{Challenge: \url{https://sites.google.com/ncsu.edu/csedm-dc-2021/}. The dataset is called ``CodeWorkout data Spring 2019'' in Datashop (\url{pslcdatashop.web.cmu.edu}).} To our knowledge, this is the only college-level, publicly-available dataset {\em with students' actual code submissions}; a concurrent work \citep{adish} uses the Hour of Code dataset, which has some similarities with this dataset but only has two questions. The CSEDM dataset contains 246 college students' 46,825 full submissions on each of the 50 programming questions over the course of an entire semester. 
%Instead of predicting the correctness of the last 20 problems given the previous answers for a student, which is the goal of the CSEDM Data Challenge, we repurpose this dataset for our task of open-ended knowledge tracing. %For each student, we would like to predict the student's next {\em full code solution}, instead of its correctness, given the student's history of solving the programming assignments. 
The dataset contains rich textual information on question prompt and students code submissions as well as other relevant metadata such as the programming concepts involved in each question and all error messages returned by the compiler. See Section~\ref{supp:exp} in the Supplementary Material for detailed data statistics and preprocessing steps. 

%truncate the sequence to length of 200 and treat the additional ones as a new student. 
% \zw{TODO: add stuff on comparisons to other datasets; why we use only one, why this one is small etc.}

%\vspace{5pt}
\noindent\textbf{Evaluation Metrics:}~
% Evaluating generative models is generally a challenging problem. 
In the context of predicting students code submissions, we need a variety of different metrics to fully understand the effectiveness of OKT. We thus use two types of evaluation metrics. First, we need metrics that can measure OKT's ability to predict student code on the test set after training. For this purpose, we use two metrics, including {\bf CodeBLEU} \citep{2020arXiv200910297R}, a variant of the classic BLEU metric adapted to code that measures the similarity between predicted code and actual student code. The other metric is the average {\bf test loss} across code tokens computed using OKT methods with the lowest validation loss. 
Second, we need metrics that can measure the diversity of predicted student code since we do not want OKT to simply memorize frequent student code in the training data. For this purpose, we use the {\bf dist-$N$} metric \citep{li-etal-2016-diversity} that computes the ratio of unique $N$-grams in the predictions over all $N$-grams. We choose $N=1$ in this work since uni-gram setting is more compatible with the limited coding vocabulary size. 
% \nl{do we need to justify this? add something but it's a strech lol} \ml{tong wen} 
We note that predicting whether a student code submission passes test cases is another important task for OKT evaluation; however, since test cases are not included in the CSEDM dataset, we cannot conduct this evaluation and leave it for future work. 

\noindent\textbf{Methods for Comparison:}~
Since exact student code prediction is a novel task, there are \textbf{no existing baselines} that we can compare against. We thus compare among variants of OKT to demonstrate that it is highly flexible and extensible. %These features enable one to plug-and-play with a variety of existing methods for the different components in the OKT framework. 
%In this work, we primarily investigate variations of the knowledge updating component in OKT which lies at the core of the OKT framework and connects to the knowledge representation and response generation modules. For this comparison, as detailed above, 
%We study the variation on several different aspects of OKT:
First, we test three different existing binary-valued KT methods, DKT, DKVMN, and AKT, as the KE component; one can apply any existing binary-valued KT method as the KE component that is suitable. As a strong baseline, we also test a version of OKT without KE and use the question prompt and code embeddings from the previous time step as additional input to the text-to-code RG component.
%We report the results on KT model variations in Section~\ref{sec:quant-results}. 
Second, we compare different alignment functions between the knowledge and question prompt embedding spaces listed in Section~\ref{sec:gen-model}. 
%several different approaches for our key technical challenge: using knowledge states as the control for the code generation model in a variety of ways.
Third, we compare several training settings, including pre-training the KE and RG components and using a multi-task training objective by adding the binary-valued response correctness prediction loss to the code generation loss in Eq.~\ref{eq:loss-gpt}, following \citep{an2022no}. 

%hether we train with the language modeling loss only (single-task) or together with the binary response correctness prediction loss (multi-task).
% \ml{discuss the pre-training/multitask things here as well?}
%training setups such as different methods to combine knowledge states computed by the KT model with the prompt for the answer generation module. 
%We report the results of the above design choices as an ablation study in Section~\ref{sec:ablation}.

\vspace{5pt}
\noindent\textbf{Experimental Setup:}
Following typical settings in the KT literature, our goal is to predict the code a student submits to a question at the next time step~$t$, $\rvx_{t+1}$, given their question prompts and code submissions in all previous time steps, i.e., $(\rvp_1,\rvx_1), \ldots, (\rvp_{t},\rvx_{t})$. 
We use two experimental settings in our experiments that capture different aspects of OKT: 
First, we analyze only the first submission to each question, ignoring later attempts. In this setting, knowledge states mostly capture a student's overall mastery of programming concepts.
Second, we analyze all code submissions from each student, including multiple consecutive attempts at the same question. In this setting, knowledge states capture not only a student's programming concept mastery but also their debugging skills. 
%This setting is used by most existing KT methods and traces students' knowledge evolution while they gradually build up their code. Therefore, This setting is also used in the literature to predict correctness on first attempt \citep{clicktrace} where consecutive time steps correspond to different questions.  
We choose not to study only the final attempt since most students were able to submit correct code in the end. See Section~\ref{supp:exp-details} of the Supplementary Material for detailed experimental settings. 
Additionally, we perform another experiment on predicting student code submissions to new questions that are unseen during training; see Section~\ref{supp:gen} for details. 
%We include additional details on experiment setup and model training in Section~\ref{supp:exp-details} of the Supplementary Material.

\begin{table}[]
%\vspace{-5pt}
\begin{adjustbox}{max width=\linewidth}
\begin{tabular}{@{}llccc@{}}
\toprule
              {\bf setting}         &   {\bf KT model}    & {\bf CodeBLEU $\uparrow$}             & \multicolumn{1}{c}{\bf Dist-1 $\uparrow$}       & {\bf Test Loss $\downarrow$}             \\ \midrule
\multirow{3}{*}{\bf first submission} & {\bf DKT}   & 0.690 &	0.422	& 0.178 \\
                       & {\bf AKT}   & 0.581 &	0.401 &	0.193 \\
                       & {\bf DKVMN} & 0.580	& 0.388 &	0.196 \\
                       & {\bf None}  & 0.518 &	0.426 &	0.215  \\
                       \midrule
\multirow{3}{*}{\bf all submissions}   & {\bf DKT}   & 0.726 &	0.403 &	0.111              \\
                       & {\bf AKT}   & 0.632 &	0.396 &	0.125            \\
                       & {\bf DKVMN} & 0.570 &	0.399 &	0.135 \\
                       & {\bf None} & 0.471 & 0.385 & 0.151 \\ 
                        \bottomrule
\end{tabular}
\vspace{2.5pt}
\end{adjustbox}
\caption{\small OKT results comparing different KT models as the KE component of OKT. AKT slightly outperforms DKVMN while DKT performs best under both settings.}
\label{tab:main1}
\vspace{-15pt}
\end{table}

\iffalse
\begin{table}[]
%\vspace{-5pt}
\begin{adjustbox}{max width=\linewidth}
\begin{tabular}{@{}llccc@{}}
\toprule
              {\bf setting}         &   {\bf KT model}    & {\bf CodeBLEU $\uparrow$}             & \multicolumn{1}{c}{\bf Dist-1 $\uparrow$}       & {\bf Test Loss $\downarrow$}             \\ \midrule
\multirow{3}{*}{\bf first submission} & {\bf DKT}   & 0.563             & 0.388 & 0.215             \\
                       & {\bf AKT}   & 0.576 & 0.395                                 & 0.200\\
                       & {\bf DKVMN} & 0.577 & 0.395                                 & 0.202 \\
                       %& {\bf None}  & 0.505 & 0.411 & 0.286 \\
                       \midrule
\multirow{3}{*}{\bf all submissions}   & {\bf DKT}   & 0.545             & 0.388                                 & 0.137              \\
                       & {\bf AKT}   & 0.565             & 0.396                                 & 0.133             \\
                       & {\bf DKVMN} & 0.519 &  0.388                                & 0.145 \\
                       %& {\bf None} & 0.456 & 0.389 & 0.249 \\
                       \bottomrule
\end{tabular}
\end{adjustbox}
\vspace{2.5pt}
\caption{Student code prediction performance for OKT on the CSEDM dataset. Using DKT, AKT, and DKVMN as the KE component result in similar performance, with AKT slightly outperforming the other two.}
\label{tab:main1}
\vspace{-15pt}
\end{table}
\fi

\subsection{Quantitative Results}
%\zw{read, discuss and restructure}
\label{sec:quant-results}
Table~\ref{tab:main1} shows the quantitative results evaluating OKT on the CSEDM dataset comparing DKT, AKT, and DKVMN as the KE component, averaged over all students and time steps. Overall, we observe that our initial OKT method performs reasonably well; as a reference, the CodeBLEU value for the examples in Table~\ref{tab:generated} are 0.8 and 0.65, respectively. Using existing binary-valued KT methods as the KE component significantly outperforms the baseline that relies on a standard text-to-code generation pipeline without this component, which suggests that KT is a key component in student-generated code prediction. Across the two experimental settings, analyzing first submissions leads to higher test loss and lower CodeBLEU score than analyzing all submissions, while performance on the Dist-1 metric does not vary much. These results can be explained by our observation that students rarely make substantial changes to their code across different submissions, often making minor tweaks; therefore, predicting a later code submission given the previous submissions becomes an easier task than predicting the first submission to a new question. Since these metrics are computed over all questions, we break down OKT's performance across questions in Section~\ref{supp:add-results} of the Supplementary Material; performance varies significantly across questions (between 0.55 and 0.85 on CodeBLEU). This observation suggests that there is considerable room for improvement on the task of exact student code prediction since they have many nuanced variations, which we further illustrate in the qualitative experiments below. 

We also see that using using DKT as the KE component of OKT significantly outperforms using AKT and DKVMN on all metrics in both experimental settings, while using AKT also outperforms DKVMN. These results suggest that DKT is more effective than AKT or DKVMN as the KE component of OKT, which also reported in \citep{pkt} for standard binary-valued KT on programming exercises, likely because DKT relies on a simple and robust LSTM model. In contrast, AKT and DKVMN have complicated model architectures and may require further parameter tuning and/or more training data in the context of OKT; typical binary-valued KT datasets are much larger in scale (up to $\sim$10M responses \citep{choi2020ednet}).  

%there is considerable room for improvement on the task of exact code prediction; in terms of CodeBLEU, OKT with DKT as its KT~\zw{is this the same as KE model?} model outperforms the \nl{what is this baseline?} \ml{this discussion needs to be updated to mesh with the new results} simple baseline of majority student code by about 0.08; as a reference, , which is slightly lower than the average performance of OKT. This observation suggests that despite OKT being able to capture major variations in student code, there are many nuances in student code that left to be captured. 

\begin{table}[]
%\vspace{-5pt}
\begin{adjustbox}{max width=\linewidth}
\begin{tabular}{@{}llccc@{}}
\toprule
                                &         & \multicolumn{1}{c}{\bf CodeBLEU $\uparrow$} & \multicolumn{1}{c}{\bf Dist-1 $\uparrow$} & \multicolumn{1}{c}{\bf Test Loss $\downarrow$} \\ \midrule
\multirow{4}{*}{\bf Alignment} & add     & 0.681 ± 0.003	& 0.423 ± 0.004 &	0.179 ± 0.006           \\
                                & average & 0.680 ± 0.003 &	0.425 ± 0.003	& 0.179 ± 0.006                \\
                                & weight  & 0.684 ± 0.008 &	0.422 ± 0.004	& 0.182 ± 0.007             \\
                                & linear & 0.696 ± 0.005 &	0.425 ± 0.004	& 0.178 ± 0.006
                                                 \\\midrule
\multirow{2}{*}{\bf Pre-train LSTM}     & yes     &    0.681 ± 0.003	& 0.423 ± 0.004 &	0.179 ± 0.006                    \\
                                & no      &     0.678 ± 0.003	        &      0.425 ± 0.002	                   &      0.180 ± 0.004              \\\midrule
\multirow{2}{*}{\bf Pre-train GPT}  & yes     &            0.702 ± 0.004	          &          0.423 ± 0.003     &      0.174 ± 0.003                \\
                                & no      &   0.678 ± 0.005	                   &     	0.415 ± 0.004          &      0.219 ± 0.006                \\\midrule
\multirow{2}{*}{\bf Multi-task} & yes & 0.706 ± 0.002 &	0.423 ± 0.002 &	0.362 ± 0.008 \\
& no  & 0.664 ± 0.019 &	0.426 ± 0.008 &	0.198 ± 0.009 \\
                                \bottomrule 
\end{tabular}
\end{adjustbox}
%\vspace{2.5pt}
\caption{\small Linearly combining knowledge states and the prompt token embeddings, pre-training both KE and RG components, and using a multi-task loss lead to best OKT performance.}
\vspace{-15pt}
\label{tab:main2}
\end{table}

Table~\ref{tab:main2} shows the quantitative results comparing different OKT designs and training settings with DKT as the KE component on first submissions. First, we see that aligning the knowledge state space with the prompt token embedding space with a learnable linear function is the most effective (with p-value of 0.01 for CodeBLEU), although other alignment functions are only slightly worse. Developing better alignment functions may further improve performance, which we leave for future work. 
%These comparisons encourage the exploration of more flexible alignment functions across these two spaces that could lead to improved performance. 
% Therefore, there is a need to explore other, more flexible alignment functions across these two spaces. 
Second, we see that pre-training the KE and RG components result in limited improvement in OKT's performance. This result suggests that there are significant differences between (i) the nature of the binary-valued KT task and OKT's exact code prediction task and (ii) code written by professionals and by students who are still learning programming. 
%using a pre-trained KT methods for response correctness prediction is not necessary in warm-starting OKT training. Also, we observe that our OKT model is capable of handling text-to-code fine-tuning process during model training itself, without the need of any additional pre-training on external code datasets. 
%These results suggest that our OKT training objective is effective in fine-tuning a generative model pre-trained on text~\zw{@lucy: check the terminology} for programming codes and adapting it for students' submissions. \zw{this last sent might not be needed}
% The unusually high diversity scores when the KE and RG components are not pre-trained is due to their inability to generate coherent code, as reflected in their low CodeBLEU scores. 
Third, we see that a multi-task OKT training objective improves both code prediction performance and model robustness in our experiments. (p-value 0.018)
This result suggests that multi-task learning with multiple objectives helps us learn better representations of the data, i.e., student knowledge state representations, in OKT. 
%Also, from our observation, adding multi-tasks help the model become more robust against other OKT settings and learning rates.

%\ml{here i think we need to discuss what these numbers mean - how good is a codebleu of 0.55? what about dist-1 of 0.38? it would be great if we can somehow relate this to the case study examples: show the codebleu for lucy's examples (prediction vs. actual) + mention roughly how many clusters are there in the big visualization plot and relate that to dist-1? we sort of need this to give people an intuitve feeling of how well our prediction works}

%We also conduct a preliminary experiment to examine OKT's performance on the standard KT prediction task, i.e., predicting the binary-valued correctness of the next response. We observe that OKT, with its additional code prediction loss in the objective, achieves a correctness prediction AUC (area under the ROC curve) value of 0.825; without this loss, the AUC value drops to 0.821, although there is no statistically significant difference. This comparison suggests that OKT has the potential to benefit the standard KT task of response correctness prediction. We leave the detailed investigation of how to use OKT to improve response correctness prediction performance as future work. 

%\vspace{-3pt}
\subsection{Interpreting Learned Knowledge States}
\label{sec:case-1}
\begin{figure*}[t]
    \centering
    \includegraphics[width=\linewidth]{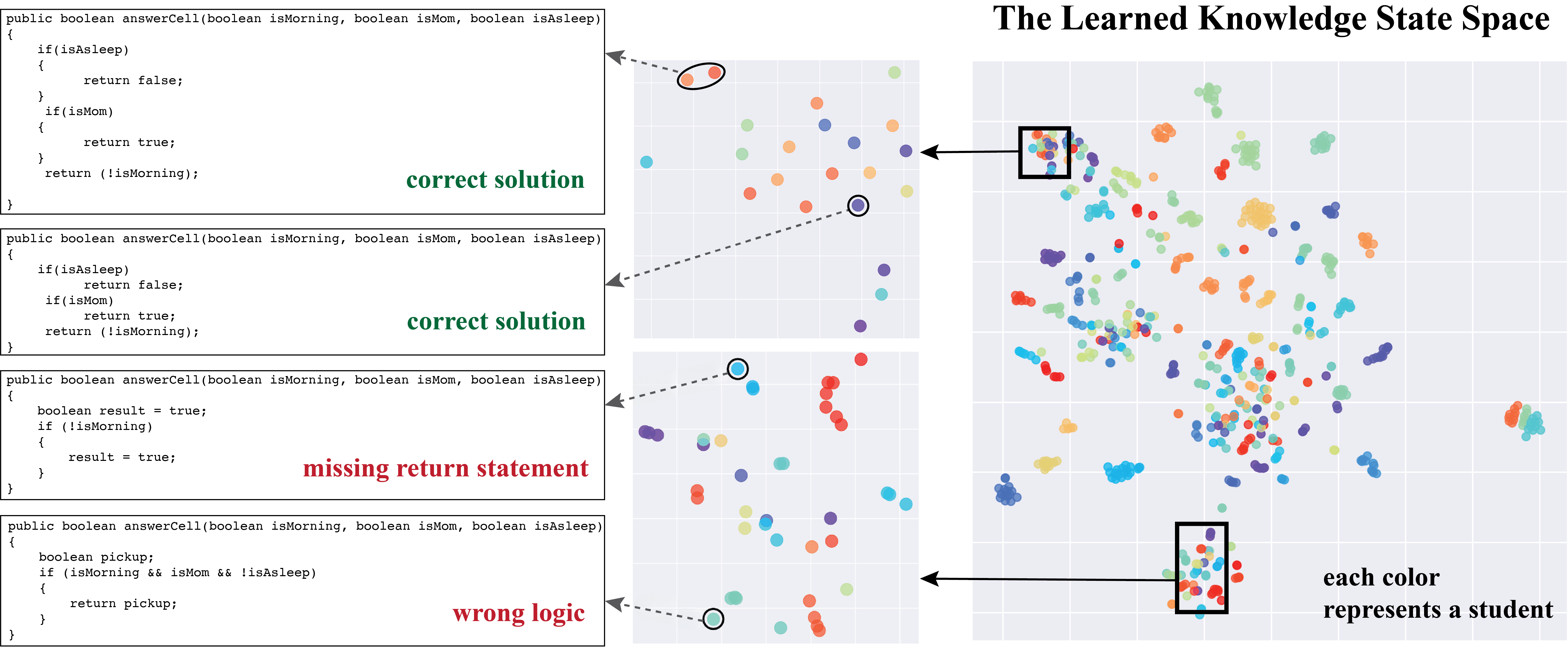}
    \vspace{-15pt}
    \caption{Visualization of latent student knowledge states (best viewed in color; each color corresponds to one student) and corresponding actual code. Knowledge states reflect the variation in student-generated code. 
    %Top Middle: visualization of the corresponding knowledge states for all student submissions. 
    %We see some well-separated clusters among student knowledge states. 
    %Top Left: zoomed-in views of two clusters and the corresponding student code where student codes in each cluster are similar. 
    %One cluster (top) corresponds to concise implementations while the other (bottom) corresponds to troubles with returning the right variable. 
    % \zw{reorganize fig2 and fig5}
    } 
    \vspace{-7pt}
    \label{fig:vis_tsne}
\end{figure*}

We now use a case study to show that the knowledge state space learned by OKT captures the variation in the content and structure of student code.  
%and effectively trace students' multiple attempts when responding to each question. 
Figure~\ref{fig:vis_tsne} visualizes the learned knowledge states, projected to a 2-D space via t-SNE \citep{tsne}, for the following question:

%\vspace{5pt}
{\small \texttt{Write a function in Java that implements the following logic: Your cell phone rings. Return true if you should answer it. Normally you answer, except in the morning you only answer if it is your mom calling. In all cases, if you are asleep, you do not answer.}}
%\vspace{5pt}

The right part of Figure~\ref{fig:vis_tsne} shows the knowledge states of all students when they respond to this question, where each dot represents the submission at a time step (a student may have multiple submissions at multiple time steps) and each color represents a student. 
We see that there are distinct clusters in these knowledge states that correspond to different student code. 
To further demonstrate this observation, we zoom in into two areas in the knowledge state space, shown in the two small plots on the left part of Figure~\ref{fig:vis_tsne} together with the corresponding actual student code submissions. 
We clearly see that the codes within each cluster share similar structural and syntactic properties and that codes from different clusters differ significantly. 
See Section~\ref{supp:rev} for a case study on how OKT's knowledge state space captures student code revisions across multiple submissions. These results suggest that the OKT-learned knowledge state space {\em aligns} with actual student code submissions. 

\begin{figure}[t!]
    \centering
    % \vspace{10pt}
    \hspace{-.3cm}
    \includegraphics[width=0.35\linewidth]{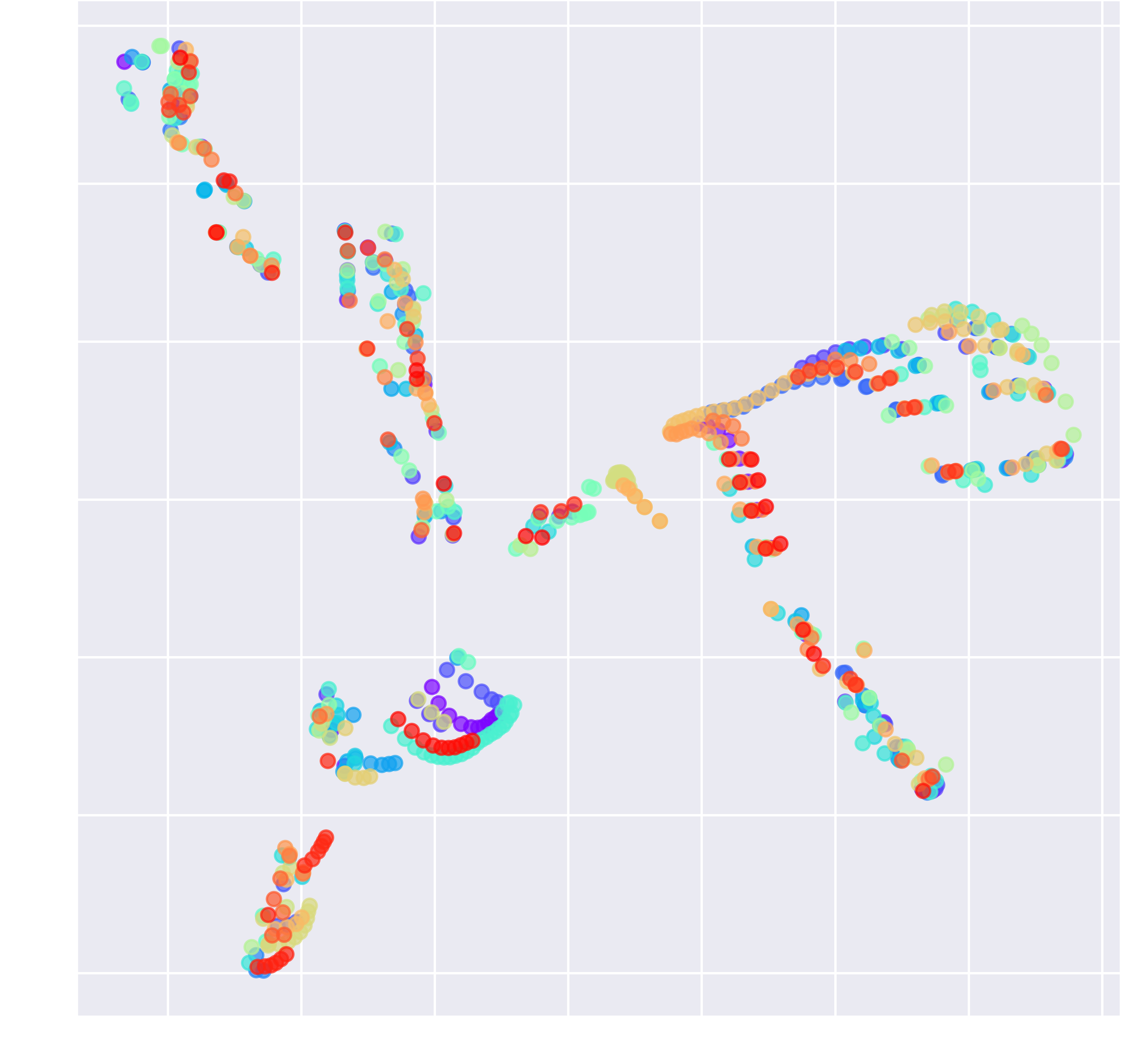} \hspace{-.3cm}
    \includegraphics[width=0.35\linewidth]{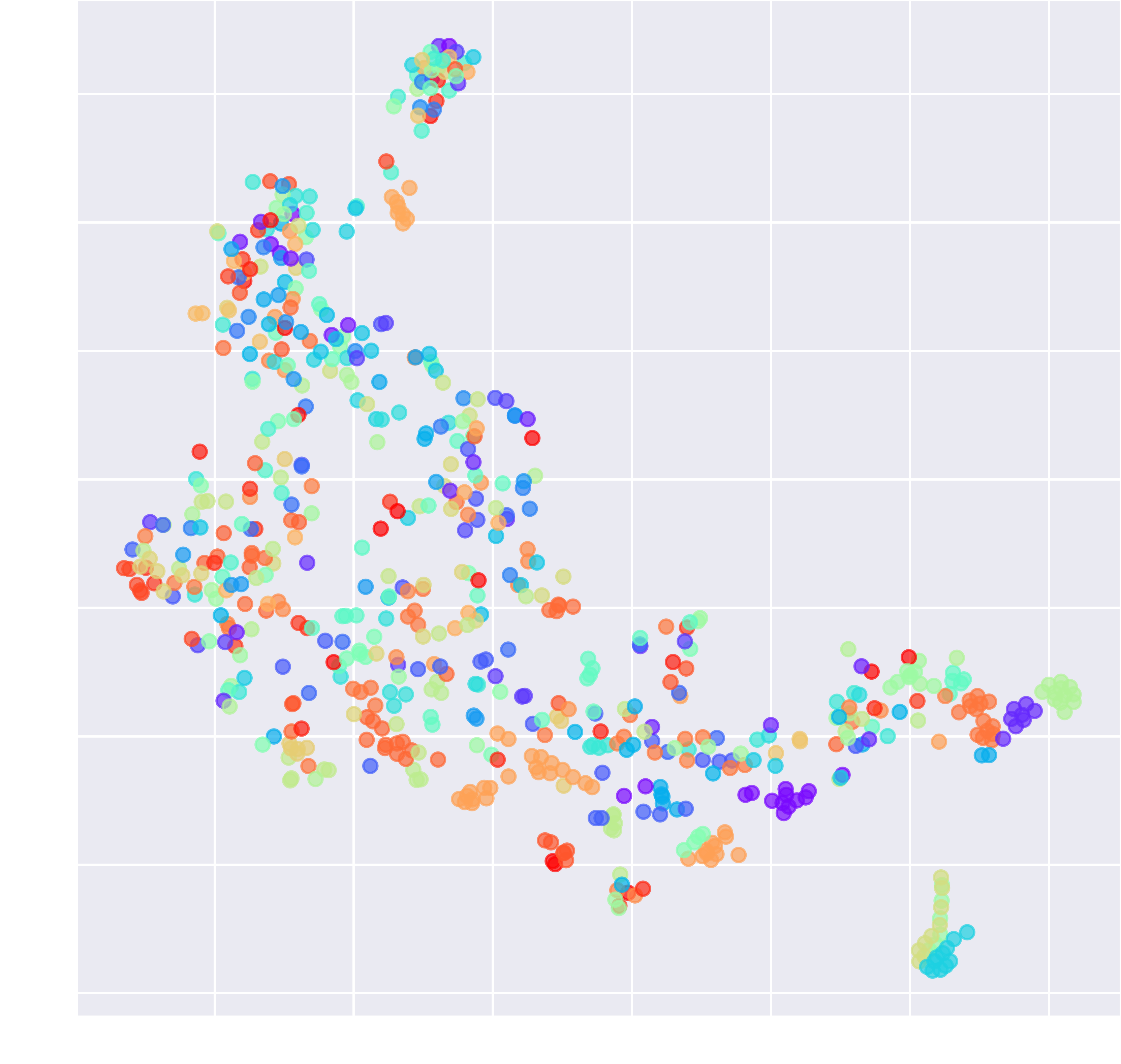} \hspace{-.3cm}
    \includegraphics[width=0.35\linewidth]{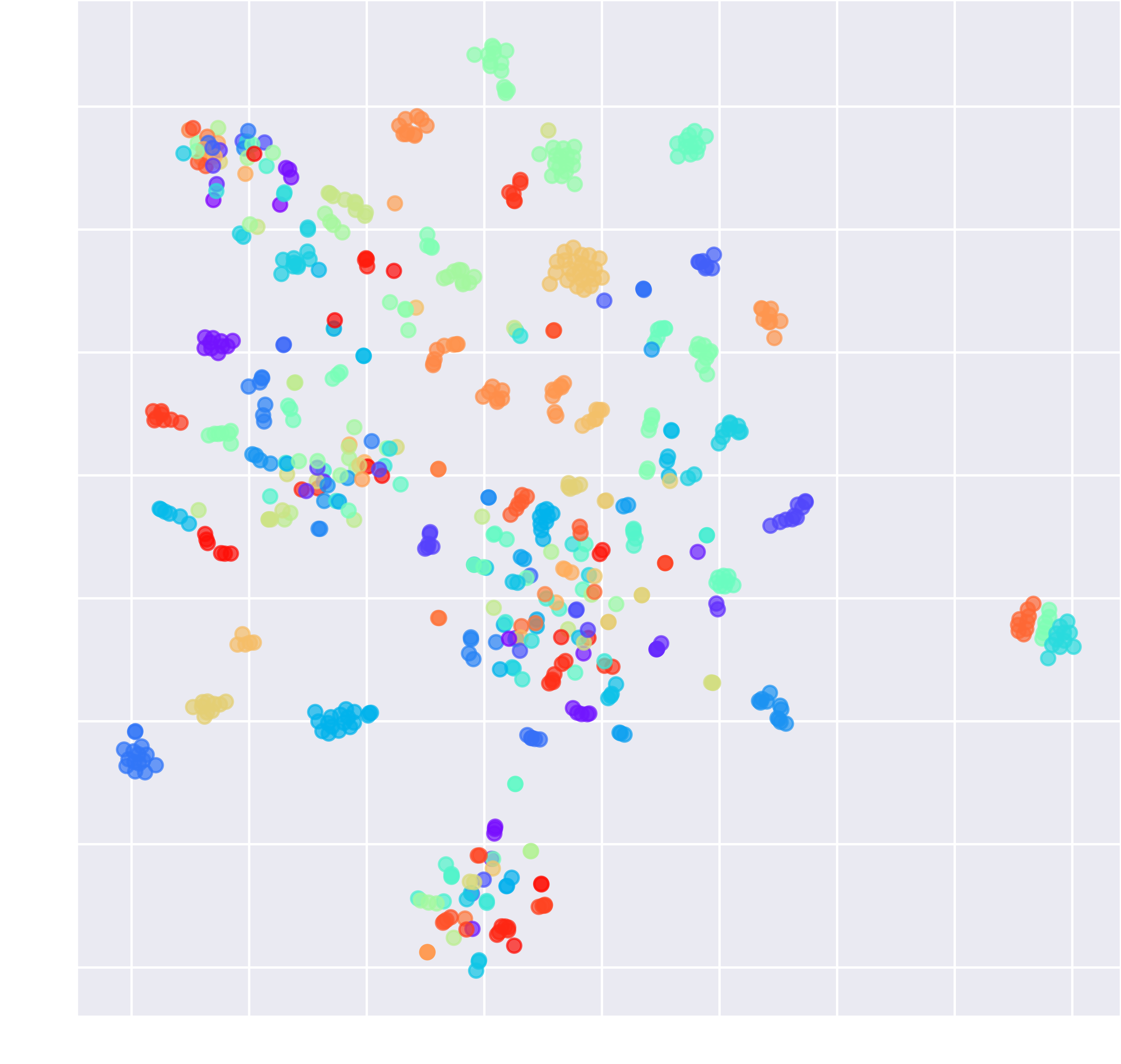}
    \vspace{-11.5pt}
    \caption{Comparison of the knowledge state spaces learned by DKT (left), DKT with code embeddings as input \citep{ncsupkt} (middle), and OKT (right). OKT learns a knowledge space with distinct clusters that capture variations in actual student code.}
    %\caption{Three different visualizations of the knowledge state spaces learnt by DKT ({\bf left}), temporal-ASTNN ({\bf middle}), and OKT ({\bf right}). The former two KT model are trained via the binary answer correctness objective whereas the our OKT model is trained via the OKT objective. The knowledge state space learnt by our OKT model is more organized and structured than the two existing KT models.}
    \label{fig:my_label}
    \vspace{-10pt}
\end{figure}

%, an indication that the learned knowledge state space captures critical information in students' codes. 
In Figure~\ref{fig:my_label}, we compare the learned knowledge state space for OKT against that for existing KT methods. We see that binary-valued DKT learns knowledge states that belong to a few highly overlapping groups with little difference within each group. The KT method in \citep{ncsupkt} that uses code embeddings only as \emph{input} to binary-valued KT learns a slightly more disentangled knowledge state space. In contrast, OKT's knowledge state space is highly informative with obvious clusters that correspond to actual student code. 
%We emphasize that existing KT methods that only analyze correctness of student responses do not capture such variation in student code since all student codes are mapped to two categories: correct and incorrect. 
%Detailed content and structural information in exact student code is lost as a result of this lossy mapping. 
% Critically, only OKT is capable of learning such space because, unlike traditional KT models that only learns to predict answer correctness, OKT is optimized on students' actual solutions, which explicitly guides the OKT model to learn knowledge states that are aware of the answer characteristics.
Overall, these results demonstrate that the knowledge state space learned by OKT captures important aspects of programming knowledge for each student. Therefore, OKT has potential in student and instructor-facing tasks such as hint generation and predicting when a student gets stuck and needs help. We can use OKT in a human-in-the-loop process for student modeling: First, OKT can identify clusters among student responses in an \emph{unsupervised} way. Then, instructors and domain experts can \emph{supervise} OKT by providing fine-grained concept or error labels on these clusters to further interpret the latent knowledge state space. 

\subsection{Knowledge-aware Prediction of Students' Code Submissions}
We now use a case study to demonstrate OKT's ability to predict student-submitted code. Similar to most existing text-to-code models \citep{concode,codexglue}, exact prediction of the actual student code is very difficult. 
However, OKT can still be effective in capturing coding styles and even predicting some error types with the help of the learned knowledge states. 
Table~\ref{tab:generated} shows the predicted code vs.\ actual student code for two questions.
For the top example, we see that our generation model is able to predict the student's code structure, capturing their use of \textit{for} loops (instead of another popular choice of \textit{while} loops). 
%The middle example shows that even though calling the \textit{concat} function is not commonly found in students' responses to this question (only 1\% of the submissions do so), OKT is still able to predict it for the student. 
In the bottom example, we see that while code prediction for this question is less accurate than for the first question, OKT can still capture the main logic and most important parts of the student's actual code. 
%This example suggests that OKT can capture both code structures and possible gaps in knowledge on programming concepts for individual students, while existing KT methods that only analyze response correctness cannot. 
%From this example, it is clear how the student makes a mistake in this question and OKT is able to successfully predict that. 
%These results show that OKT can potentially be used to provide automated feedback to both students and instructors. 
These examples show that OKT can capture both code structure and knowledge gaps on programming concepts for individual students and even predict their possible errors; this capability has much more potential for student and instructor support than standard binary-valued KT methods.

\begin{table}[t] 
\centering 
\begin{adjustbox}{max width=1\linewidth}
\begin{tabular}{|p{4cm}|p{4cm}|}       
\hline                                   
\multicolumn{1}{|c|}{\textbf{predicted code}}  & \multicolumn{1}{|c|}{\textbf{actual student code}} \\ 
\hline
\includegraphics[height=0.94in]{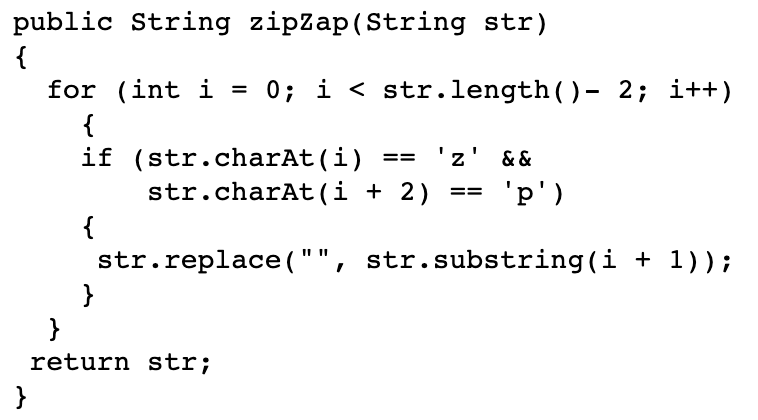}& \includegraphics[height=0.99in]{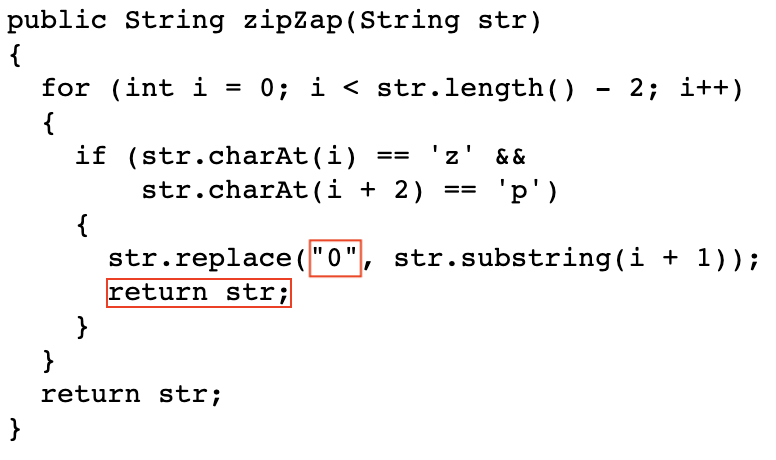} \\
%\hline  
%\includegraphics[height=1.8in]{figures/gen_2a.png}& \includegraphics[height=1.8in]{figures/gen_2b.png} \\
\hline  
\includegraphics[height=0.933in]{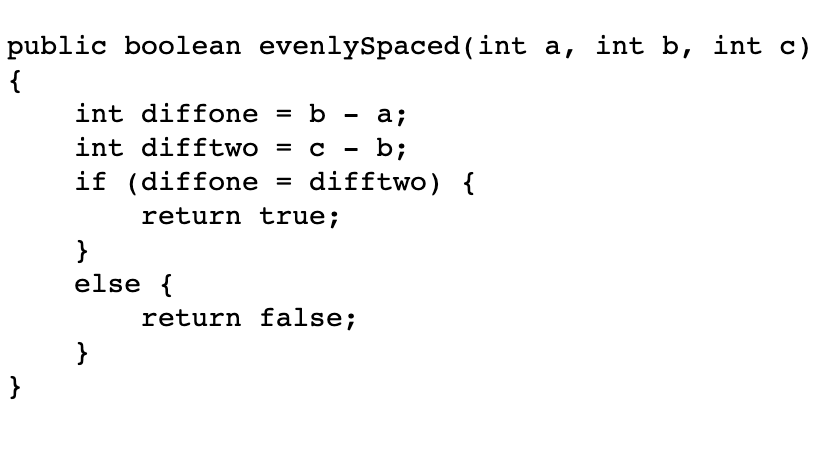}& \includegraphics[height=0.9in]{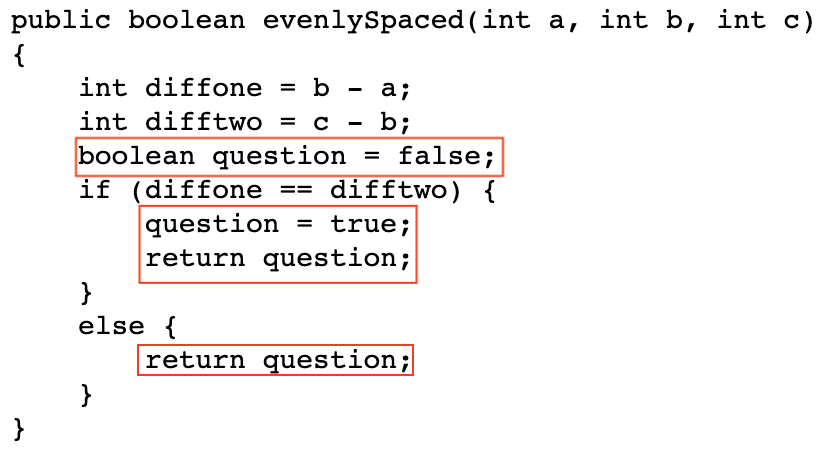} \\
\hline  
\end{tabular}
\end{adjustbox}
%\vspace{5pt}
\caption{OKT generated code vs.\ actual student code for two questions (differences highlighted in red boxes).}
\vspace{-5pt}
\label{tab:generated}
\end{table} 

\subsection{Generalizing to Unseen Questions}
\label{supp:gen}

One important limitation of binary-valued KT methods is that they cannot really generalize to new questions; if a question is not present during training, these methods can only predict a student's probability of responding to it correctly using its concept labels (which are often unavailable). On the contrary, OKT's KR and RG components utilize exact question and response content, enabling it to generalize to new questions and predict exact responses to these questions and specific errors. We conduct a preliminary experiment to demonstrate this advantage of OKT: we first remove one question from the dataset (say it occurs at time step $t$ for a student) and then 
%, namely, $q_t$, from the dataset as the unseen question. Then we 
predict the response to this question using the estimated knowledge state $\rvh_t$. 
We explore two ways to estimate $h_t$: i) averaging the knowledge states from neighboring time steps, i.e., $\rvh_{t-1}$ and $\rvh_{t+1}$, 
%from its neighboring questions $q_{t-1}$, $q_{t+1}$ corresponding.
and ii) using the knowledge state from the previous time step, i.e., $\rvh_{t-1}$. 
As a baseline, we also use randomly generated knowledge state vectors to predict the response. 
%(2). Use average of the knowledge states from previous and next questions, $h_t = \frac{h_{t-1} + h_{t+1}}{2}$. We also use random numbers from the same range as $h_t$ as the baseline to compare with. 

\begin{table}[t]
\begin{adjustbox}{max width=\linewidth, center}
\begin{tabular}{@{}lcc@{}}
% \vspace{3pt}
\toprule
\textbf{Method} & \textbf{CodeBLEU $\uparrow$} & \textbf{Dist-1 $\uparrow$} \\ \hline
Previous   &      0.484     &     0.431     \\
Average     &    0.504     &         	0.419        \\
Random  &          0.328   &    0.452              \\  \bottomrule 
\end{tabular}
\end{adjustbox}
\caption{OKT's generalization performance to new questions that are unseen during training, using knowledge states from the previous time step, neighboring time steps, and random values.}
\vspace{-13pt}
\label{tab:remove_res}
\end{table}

\begin{figure}[ht!]
    \centering
    % \vspace{-5pt}
    % \hspace{-0.2cm}
    \includegraphics[width=.9\linewidth]{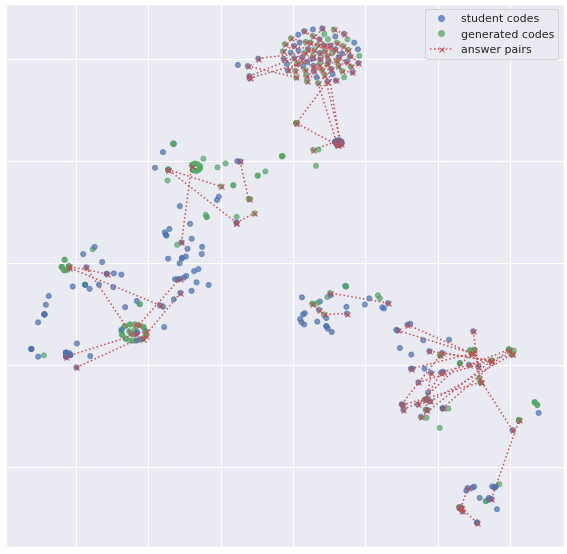} 
    \vspace{-5pt}
    \caption{Visualization of actual student code (blue) compared to predicted code (green) for a new question unseen during training. Code pairs that are close in the code embedding space are connected (red).}
    \label{fig:remove_Q}
    \vspace{-7pt}
\end{figure}

Table~\ref{tab:remove_res} shows the average results over removing each question, using DKT on first submissions. 
We use a smaller amount of epochs for this experiment (10 compared to 25 epochs from Table~\ref{tab:main1}), which explains some of the significant drop in CodeBLEU scores. Nevertheless, OKT still significantly outperforms the baseline approach with no KE component, with averaging knowledge states from neighboring time steps slightly outperforming using the previous time step. 
% We see that despite the CodeBLEU numbers dropping significantly compared to Table~\ref{tab:main1}, we still blahblahblah. 
Figure~\ref{fig:remove_Q} visualizes predicted code vs.\ actual student code embeddings for an unseen question with an average CodeBLEU value of 0.538 over all students. Blue dots correspond to actual student responses and green dots represent RG predicted responses in 2-D, while red dots correspond to pairs of predicted and actual code that are highly similar (76 out of 225). 
%We also examine the distance between student written and generated codes and connect the relatively close ones with red dotted lines. For this question, we have 225 pairs of student written and generated codes with an average CodeBLEU score of 0.538, where 76 pairs with CodeBLEU of 0.729 are selected. 
We clearly see that OKT is able to capture the majority of student code variations on this new question from their responses to other questions and left no parts of the code embedding space unaccounted for. 
%there exists an oval-shaped area that contains many closely linked response pairs. Also, the connected pairs in the bottom left part shows that despite not observing the question during training process, OKT is still able to predict answers written in a different code style. 
OKT's capability of generalizing to new questions can potentially be used to provide feedback to teachers plan homeworks, by predicting typical errors in programming questions that students in their class may exhibit, before assigning them.

\section{Related Work}
\label{sec:rw}

\noindent{\bf Knowledge Tracing:}~
Existing methods for binary-valued KT can be broadly grouped by how they represent the student knowledge level variable, $\rvh$, in Eq.~\ref{eq:ku}. For example, classic Bayesian knowledge tracing methods \citep{mozerfuse,ktcomparepardos,yudelson} treat student knowledge as a binary-valued latent variable. The KE and RP components are noisy binary channels, resulting in excellent interpretability of the model parameters. 
Factor analysis-based methods \citep{lfa,das3h,pfa} use features and latent ability parameters to model student knowledge. The RP component in these methods relies on item response theory models \citep{irtbook}. 
More recently, deep learning-based KT methods \citep{akt,rkt,dkt,saint+,dkvmn} treat student knowledge as hidden states in neural networks. The KE component often relies on variants of recurrent neural networks \citep{lstm}, resulting in models that excel at future performance prediction but have limited interpretability.
% but often limited interpretability.

Student responses, i.e., $\rvx$ in Eq.~\ref{eq:ku}, are almost always treated as a binary-valued scalar indicating response correctness. Few methods characterize them as non-binary-valued such as option tracing \citep{ot}, which analyzes the exact option students select on each multiple-choice question, and predict partial analysis \citep{neilkt}. 
%In general, one can use polytomous IRT models \citep{polytomous} as the RP component in KT methods to predict categorical-valued (such as options in multiple-choice questions) and ordinal-valued (such as partial credit) responses \citep{ncdm,sparfatag}. 
Questions, i.e., $\rvq$ in Eq.~\ref{eq:ku}, are often one-hot encoded, either according to question IDs/concept tags, or in a few cases, represented with graph neural networks using question-concept dependencies \citep{gikt}. Few existing works use exact question content for $\rvq$. For example, \citep{eernna,purdue} use pre-trained word embeddings such as word2vec \citep{word2vec} to encode questions in the RP component. Specifically for programming questions, \citep{peach,ncsupkt,pkt} use code representation techniques such as ASTNN \citep{astnn} and code2vec \citep{code2vec} to convert student code into vectors and use them as input to the KE component. 

%\vspace{-5pt}
\noindent{\bf Program Synthesis and Computer Science Education:}~ 
Program synthesis from natural language instructions \citep{programsynthesis} has attracted significant recent interest since pre-trained language models \citep{codex} or language model architectures \citep{alphacode} have demonstrated their effectiveness on hard tasks such as solving coding challenge problems \citep{apps}. These methods are pre-trained on large datasets containing publicly available code on the internet, which is primarily written by skilled programmers. There is a line of existing work on analyzing student-generated code, most noticeably using the Hour of Code dataset released by Code.org \citep{peaches,adish,peach}, for tasks such as error analysis and automated feedback generation that are meaningful in computer science education settings.

\section{Conclusions and Future Work}

In this paper, we have proposed a framework for open-ended knowledge tracing (OKT) to track student knowledge acquisition while predicting their full responses to open-ended questions. We have demonstrated how OKT can be applied to the computer science education domain, where we analyze students' code submissions to programming questions. We addressed the key technical challenge of integrating student knowledge representations into code generation methods, e.g., text-to-code models based on fine-tuning GPT-2. 
%Since exactly predicting open-ended responses, especially longer ones, can be difficult, we also defined a series of metrics to evaluate the performance of open-ended knowledge tracing methods. 
Our experiments on real-world computer science student data indicate that OKT has considerable promise for tracking and predicting student mastery and performance.

%, we discussed the effectiveness and limitations of our framework, both quantitatively and qualitatively.

There are many avenues for future work. 
%First, we need to dig deeper and find the best ways to inject knowledge states into generative models of student code, such as using prefixes \citep{prefix}. 
First, we can use code standardization techniques \citep{ken} to further pre-process student code using semantic equivalence. 
Second, we can explore the applicability of OKT to other domains such as mathematics, where many pre-trained models for mathematical problem solving have been developed \citep{verifier,math,deepmind} and explore whether students consistently exhibit certain errors \citep{kurt}. 
Third, we can develop knowledge tracing models that capture more specific aspects of knowledge, i.e., debugging skills, which is reflected in the \emph{change} in student code across submissions to the same question after receiving automated feedback generated by the compiler or test cases.
Fourth, we can further enhance the validity and interpretability of OKT by adding more human supervision, such as adding an additional loss on the test case scores of generated code. We can also use instructor- or expert-provided labels on student errors to make the latent knowledge state space more informative. 
Finally, we can further evaluate our framework on tasks relevant to instructor feedback, including compilation/runtime error category prediction and test case outcome prediction; see Section~\ref{supp:usecase} in the Supplementary Material for a detailed discussion. 

\newpage
\section*{Acknowledgements}
We thank the reviewers for their constructive feedback. This work is supported by NSF grants 1842378, 1937134, 1917713, 2118706, 2202506, 2215193, ONR grant N0014-20-1-2534, AFOSR grant FA9550-18-1-0478, ONR grant N00014-18-1-2047, and a Vannevar Bush Faculty Fellowship.

\section*{Limitations}

Being the first attempt at the task of predicting the exact content of open-ended student responses, OKT has several obvious limitations. First, the ability to predict variation in student responses depends on the fine-tuned language model's ability to generate correct responses given the question statement. Therefore, it is not clear whether OKT can generalize to domains where language models have not been shown to be highly accurate at generating correct open-ended responses. Second, OKT requires a large amount of student coding data, which may limit its applicability to learning platforms in their early stages that do not have a large number of student users. Third, the open-ended response generation process is sequential and can be time-consuming, which may limit OKT's ability to support instructors and students in real time in real-world computer science education scenarios. 
%we do not perform any code standardization since we believe that raw student code contains nuanced information about their knowledge; over-standardization loses this information. Nevertheless, some amount of standardization could be beneficial \cite{ken}

\section*{Ethics Statement}
Our work should be seen as exploratory rather than a finished tool that can readily be deployed in real-world computer science educational scenarios. Since OKT requires training on a large amount of student-generated code, there is a need to systematically study any potential negative biases towards underrepresented student populations. The effectiveness of exact open-ended response prediction in helping instructors adjust their instruction and benefit students remains to be seen, which requires principled evaluation using A/B testing.

\bibliography{custom}
\bibliographystyle{acl_natbib}

\newpage 
\appendix

\section{Dataset Statistics and Preprocessing Steps}
\label{supp:exp}

\begin{table}[h!]
\begin{adjustbox}{max width=\linewidth}
\begin{tabular}{@{}lcc@{}}
\toprule
                 \textbf{Statistic}                   & \textbf{Raw} & \textbf{Processsed} \\ \midrule
\#codes                              & 46825        & 39796               \\
\#avg. lines of code per submission & 17.52        & 17.64               \\
\#avg. submissions per student       & 190.34       & 161.77              \\
\#avg. submissions per problem       & 936.5        & 795.9               \\ \bottomrule
\end{tabular}
\end{adjustbox}
\vspace{2pt}
\caption{\small Dataset statistics comparing the raw and our processed dataset, the latter of which is used throughout our experiments.}
\label{tab:data-stat}
\vspace{-5pt}
\end{table}

Since we choose to use the AST representation for code, we perform a preprocessing step to remove student solutions that cannot be converted to AST format. Overall, about 85\% of all student solutions are AST-convertible, which means that this preprocessing step does not result in significant data loss. Table~\ref{tab:data-stat} describes the summary statistics of the original dataset and the resulting preprocessed dataset that we use for all our experiments. For KT, we follow standard procedure in the literature \citep{dkt,akt,dkvmn} by setting the maximum solution sequence for any student to 200. For students with more than 200 solutions, we split their solutions into separate sequences of length 200. 

% move this here so that it's in previous page
\begin{figure*}[h!]
    \centering
    \hspace{-0.2cm}
    \includegraphics[width=0.33\linewidth]{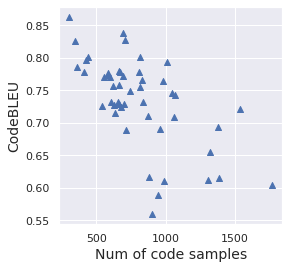} \hspace{-0.2cm}
    \includegraphics[width=0.33\linewidth]{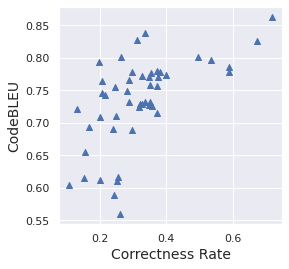} \hspace{-0.2cm}
    \includegraphics[width=0.33\linewidth]{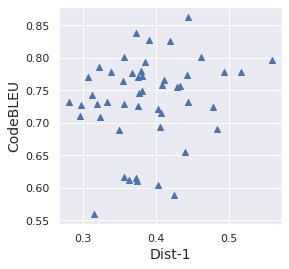}
    \vspace{-7pt}
    \caption{Visualization of CodeBLEU metric versus number of student responses (left), rate of correct submissions (middle) and Dist-1 metric (right) in each question. Each point represents one question.}
    \label{fig:codebleu}
\end{figure*}

\section{Experimental Setup Details}
\label{supp:exp-details}
%data regimes where the model receives either (1) all code solutions from each student or (2) the first code solution to each problem from each student. The former regime is useful for tracing students' multiple attempts when they solve for a single problem while the latter regime is useful for tracing students' first attempt for each problem \zw{motivation for the second thing?}. 
In both settings, we split all students in the dataset into disjoint (train, validation, test) sets with a $80\%-10\%-10\%$ ratio and report all metrics on the test set. 
In the default setting, we add knowledge states into token embeddings for the question prompt as input to the RG component using a linear projection layer as detailed above. 
We pre-train both the KE and RG components of OKT on the training data with a correctness prediction objective (i.e., pre-train an existing KT method) and a prompt-to-code supervised generation objective, respectively. For the KE component, we follow the original DKT, DKVMN, and AKT methods with $768$ hidden units in their models.
When combining DKVMN or AKT with the answer generator, we use the context reader output from DKVMN and the hidden state from AKT, respectively, as input to the answer generator at each time step. We refer readers to \citep{dkvmn,akt} for more details.
For the RG component, we use a small GPT-2 with 12 transformer decoder layers \citep{gpt-2}. We use the RMSProp optimizer for the knowledge update component and the Adam optimizer for the RG component, both with a default learning rate of $0.00001$. Also, we freeze the parameters of the question and code representation models and only train the KT model and the answer prediction model, Although the former two components can also be optimized.

We run all experiments using a single NVIDIA Quadro RTX 8000 GPU. The KT model pre-training usually takes less than 5 minutes per epoch of wall clock time. The OKT training with DKT as the KT model takes about 10 minutes and 30 minutes per epoch of wall clock time for the the two scenarios, namely, using only students' first submitted code and all submitted code that can be converted to AST format, respectively. OKT training with AKT as the KT model takes about the same time as DKT as the KT model while with DKVMN, training is about 1.5 times slower due to the more expensive memory computation \citep{dkvmn}.

%\section{OKT Framework Design Choices} 
%\label{sec:ablation}
%We conduct an ablation study to understand how each design choice in OKT impacts its performance on predicting student code submissions. 
%To efficiently evaluate a number of different configurations, we choose to model students' first code submission for each prompt in this experiment. 
%For this experiment, we analyze only the first code submission for each question made by every student. We compare four different model design aspects: First, we compare different {\bf alignment functions} between knowledge states and prompt token embeddings in the response generation component that we detailed above in Section~\ref{sec:gen-model}. Second and third, we test whether {\bf knowledge update component pre-training} and {\bf response generation component pre-training} are effective. Fourth, we compare whether using a {\bf multi-task} training objective where we add the typical correctness prediction loss to the code prediction loss in Eq.~\ref{eq:loss-gpt} during OKT training leads to improved code prediction accuracy. 

%
%verifies our intuition that OKT will perform well when training on top of existing KT methods  and answer prediction are already adapted to the given educational domain. This result also suggest that, for effective OKT, one should first adapt KT and answer predictor for the given subject domain via pre-training. 

%corroborates the results in multiple recent works that suggest multi-task as an effective approach for improving model performance. 

\begin{figure*}[ht!]
    \centering
    \includegraphics[width=\linewidth]{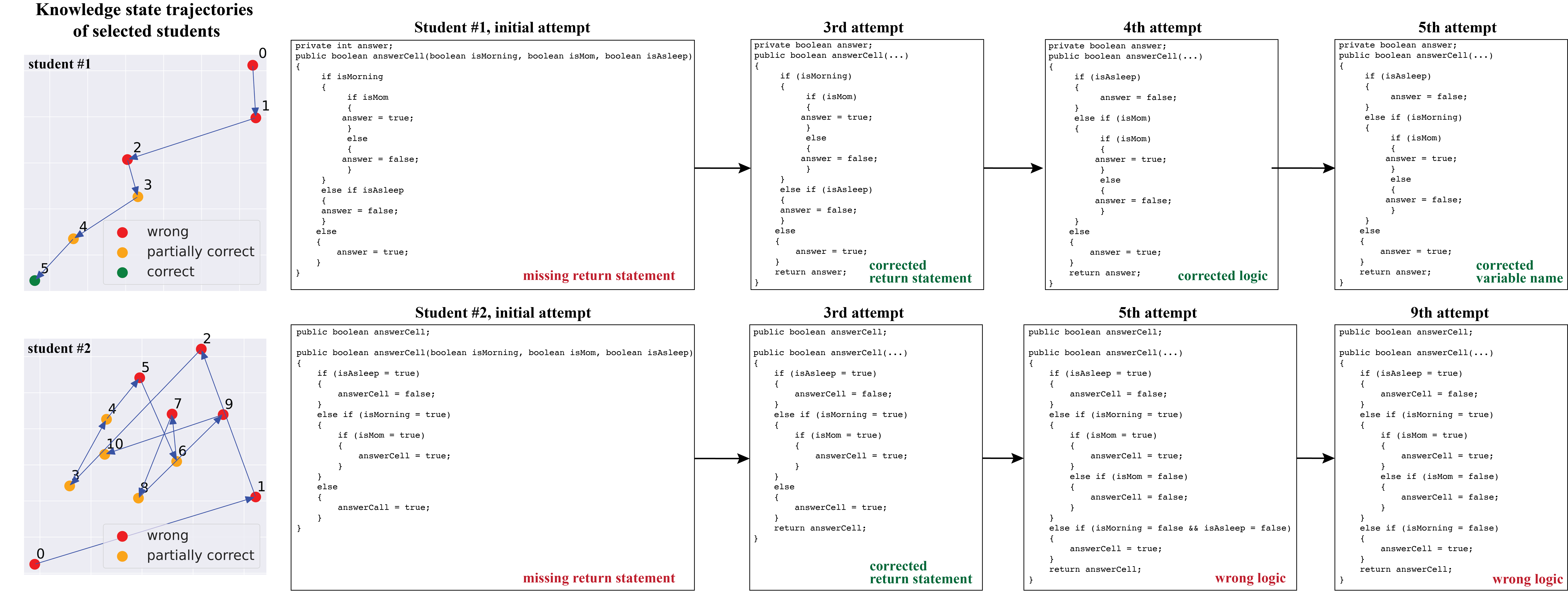}
    \vspace{-15pt}
    \caption{Four sample submissions of two students corresponding to the top right figures, respectively. One student gradually proceeded to a correct code while the other got stuck.
    } 
    \vspace{-13pt}
    \label{fig:traces}
\end{figure*}

\section{Visualizing Quantitative Results}
\label{supp:add-results}

Following the results in Section~\ref{sec:quant-results} and Table~\ref{tab:main1}, we additionally examine the model performance across questions and measure the correlation between its CodeBLEU score and some features (i.e. difficulty level, response diversity). Figure~\ref{fig:codebleu} shows that model performance has a positive correlation with student performance, i.e., the portion of correct responses, and a negative correlation with the number of student responses. In other words, an easy question with fewer submissions is more likely to achieve better prediction results. However, CodeBLEU is minimally correlated with the diversity in student responses. Also, the range of CodeBLEU performance across questions is relatively big, with the highest of 0.86 and lowest of 0.56.

\section{Visualizing Code Revisions}
\label{supp:rev}
We also show how the learnt knowledge state space can be useful for tracing and understanding students' consecutive submissions to the same question. 
On the right-hand side of Figure~\ref{fig:traces}, we show the knowledge state trajectories of two students responding to this question. 
The colors in these two figures represent knowledge states that correspond to wrong, partially correct and fully correct codes at different time steps. 
We see that both students start with a wrong solution. However, one student gradually proceeded to the correct solution after a few edits, whereas the other student got stuck after a few unsuccessful edits and eventually gave up on solving this prompt. The steady progress versus getting stuck is clearly visualized in these figures, where for the former student, the knowledge states gradually moves from the upper right corner in the knowledge state space to the lower left, whereas for the latter student, the knowledge states circle around and bounce back and forth in the space. 
We also show four selected submissions by each student during their response process, further illustrating how the first student made steady progress, i.e., adding the return statement (first two code submissions) and correcting logic (last two code submissions), and how the second student got stuck, i.e., making reasonable changes initially but then some repetitive edits.

\section{Real-World Use Cases and Implications}
\label{supp:usecase}
\vspace{-5pt}
One crucial highlight of our work is that, through OKT, we can predict students' responses to open-ended questions {\it before actually assigning them}. On the contrary, existing student and teacher support tools can only be applied {\it after} observing their responses. Therefore, OKT can be used in practice in many ways by anticipating student errors and struggles ahead of time. For example, for teacher support, we can use OKT to provide diagnosis information to teachers via a dashboard. For any open-ended question that the teacher wants to assign to their class, we can predict the responses that each student will write given their current knowledge states and show teachers clusters that represent typical errors. This way, the teacher can anticipate student performance, switch to an easier (or more challenging) question if necessary, and prepare feedback for individual students ahead of time. For student support, if a student struggles, we can use OKT to find other students stuck in a similar place but ultimately succeeded in answering the question and provide incremental hints or feedback on their errors. 
These advantages over traditional KT methods will potentially enable OKT to become the next-generation workhorse for large-scale, intelligent educational systems. 

\end{document}